%
\let\useblackboard=\iftrue
%
%
\newfam\black
\input harvmac.tex
\def\Title#1#2{\rightline{#1}
\ifx\answ\bigans\nopagenumbers\pageno0\vskip1in%
\baselineskip 15pt plus 1pt minus 1pt
\else
\def\listrefs{\footatend\vskip 1in\immediate\closeout\rfile\writestoppt
\baselineskip=14pt\centerline{{\bf References}}\bigskip{\frenchspacing%
\parindent=20pt\escapechar=` \input
refs.tmp\vfill\eject}\nonfrenchspacing}
\pageno1\vskip.8in\fi \centerline{\titlefont #2}\vskip .5in}

\ifx\answ\bigans\def\tcbreak#1{}\else\def\tcbreak#1{\cr&{#1}}\fi
\useblackboard
\message{If you do not have msbm (blackboard bold) fonts,}
\message{change the option at the top of the tex file.}
\font\blackboard=msbm10 scaled \magstep1
\font\blackboards=msbm7
\font\blackboardss=msbm5
\textfont\black=\blackboard
\scriptfont\black=\blackboards
\scriptscriptfont\black=\blackboardss

\else

\fi
%
\def\yboxit#1#2{\vbox{\hrule height #1 \hbox{\vrule width #1
\vbox{#2}\vrule width #1 }\hrule height #1 }}
\def\fillbox#1{\hbox to #1{\vbox to #1{\vfil}\hfil}}
\def\ybox{{\lower 1.3pt \yboxit{0.4pt}{\fillbox{8pt}}\hskip-0.2pt}}

\def\comments#1{}

\def\half{{1\over 2}}

\def\CM{{\cal M}}

\def\II{\relax{I\kern-.07em I}}

\def\IZ{\relax\ifmmode\mathchoice
{\hbox{\cmss Z\kern-.4em Z}}{\hbox{\cmss Z\kern-.4em Z}}
{\lower.9pt\hbox{\cmsss Z\kern-.4em Z}}
{\lower1.2pt\hbox{\cmsss Z\kern-.4em Z}}\else{\cmss Z\kern-.4em
Z}\fi}
\def\IB{\relax{\rm I\kern-.18em B}}
\def\IC{{\relax\hbox{$\inbar\kern-.3em{\rm C}$}}}
\def\ID{\relax{\rm I\kern-.18em D}}
\def\IE{\relax{\rm I\kern-.18em E}}
\def\IF{\relax{\rm I\kern-.18em F}}
\def\IG{\relax\hbox{$\inbar\kern-.3em{\rm G}$}}
\def\IGa{\relax\hbox{${\rm I}\kern-.18em\Gamma$}}
\def\IH{\relax{\rm I\kern-.18em H}}
\def\II{\relax{\rm I\kern-.18em I}}
\def\IK{\relax{\rm I\kern-.18em K}}
\def\IP{\relax{\rm I\kern-.18em P}}

\font\cmss=cmss10 \font\cmsss=cmss10 at 7pt
\def\IR{\relax{\rm I\kern-.18em R}}

\def\sdtimes{\mathbin{\hbox{\hskip2pt\vrule
height 4.1pt depth -.3pt width .25pt\hskip-2pt$\times$}}}

\def\BR{\IR}
\def\BZ{\IZ}

\def\sym#1{{{\rm SYM}} _{#1 +1}}

\def\IZ{\relax\ifmmode\mathchoice
{\hbox{\cmss Z\kern-.4em Z}}{\hbox{\cmss Z\kern-.4em Z}}
{\lower.9pt\hbox{\cmsss Z\kern-.4em Z}}
{\lower1.2pt\hbox{\cmsss Z\kern-.4em Z}}\else{\cmss Z\kern-.4em
Z}\fi}
\def\IB{\relax{\rm I\kern-.18em B}}
\def\IC{{\relax\hbox{$\inbar\kern-.3em{\rm C}$}}}
\def\ID{\relax{\rm I\kern-.18em D}}
\def\IE{\relax{\rm I\kern-.18em E}}
\def\IF{\relax{\rm I\kern-.18em F}}
\def\IG{\relax\hbox{$\inbar\kern-.3em{\rm G}$}}
\def\IGa{\relax\hbox{${\rm I}\kern-.18em\Gamma$}}
\def\IH{\relax{\rm I\kern-.18em H}}
\def\II{\relax{\rm I\kern-.18em I}}
\def\IK{\relax{\rm I\kern-.18em K}}
\def\IP{\relax{\rm I\kern-.18em P}}

\font\cmss=cmss10 \font\cmsss=cmss10 at 7pt
\def\IR{\relax{\rm I\kern-.18em R}}

\def\sdtimes{\mathbin{\hbox{\hskip2pt\vrule
height 4.1pt depth -.3pt width .25pt\hskip-2pt$\times$}}}

\def\bT{{\bf T}}
\def\bS{{\bf S}}

\def\tilde{\widetilde}
\Title{ \vbox{\baselineskip12pt\hbox{hep-th/9702187}
\hbox{RU-97-6}}}
{\vbox{\centerline{Strings from Matrices${}^*$}}}

{}\footnote{}{${}^*$ A preliminary version of this paper was presented
at a lecture on January 5, 1997 at the Jerusalem Winter School on
Strings and Duality.}

\centerline{Tom Banks and Nathan Seiberg }
\smallskip
\smallskip
\centerline{Department of Physics and Astronomy}
\centerline{Rutgers University }
\centerline{Piscataway, NJ 08855-0849}
\centerline{\tt banks, seiberg @physics.rutgers.edu}
\bigskip
\bigskip
\noindent
We identify Type IIA and IIB strings, as excitations in the matrix model
of M theory. 

\Date{December 1996}

\newsec{\bf Introduction}

\nref\dhn{B. de Wit, J. Hoppe and H. Nicolai, {\it On the Quantum
Mechanics of Supermembranes}, Nucl.Phys. {\bf B305[FS23]}(1988)545.}%
\nref\towns{P.K. Townsend, {\it D-branes {}from M-branes}, Phys. Lett.
{\bf B373} (1996) 68.}%
\nref\bfss{T. Banks, W. Fischler, S.H. Shenker and L. Susskind, {\it M
Theory as a Matrix Model: a Conjecture}, hep-th/9610043.}%

The matrix model of M theory \refs{\dhn - \bfss} purports to be a
unified description of all string vacuum states, as well as a
nonperturbative quantum mechanical framework for string theory.
In this paper we will demonstrate the existence of some of the degrees
of freedom that describe these different vacuum states.  We will find
freely propagating ten dimensional Type IIA and Type IIB strings.
Results related to ours have recently been
obtained by Sethi and Susskind
\ref\ss{S. Sethi, L.Susskind, {\it Rotational Invariance in the M(atrix)
Formulation of Type IIB Theory}, hep-th/9702101. }
and by Motl
\ref\Motl{L. Motl, {\it Proposals on Nonperturbative Superstring
Interactions}, hep-th/9701025.}.

We begin by recalling and refining the description of Type IIA strings
\bfss\ in the matrix model.  One compactifies the ninth transverse
dimension of the matrix model on a circle of radius $R_9$ by restricting
attention to large $N$ matrices of the form
\eqn\compacta{X^9 = {1\over i}{\partial\over \partial\sigma} -
A(\sigma)}
$A$ is a $U(M)$ gauge potential, and $M$ goes to infinity.  
The other matrix degrees of freedom are restricted to be functions of
$\sigma$ which transform in the adjoint of $U(M)$.  $\sigma$ has period
$1/{R_9}$.  This ansatz
\nref\taylor{W. Taylor, {\it D-brane Field Theory on Compact Spaces},
hep-th/9611042.}%
\nref\compact{L. Susskind, {\it T Duality in M(atrix) Theory and S
Duality in Field Theory}, hep-th/9611164.}%
\nref\princeton{O. Ganor, S. Rangoolam and W. Taylor,{\it Branes, Fluxes
and Duality in M(atrix) Theory}, hep-th/9611202.}%
\refs{\bfss,\taylor,\compact,\princeton}
is motivated by the observation that shifts of $X^9$ by $2\pi R_9$ are
gauge transformations (in the full matrix model gauge group, not just its
$U(M)$ subgroup) and by extrapolating the description of zero branes in
weakly coupled IIA string theory.  For higher dimensional tori, one
obtains, by analogous arguments, the dimensional reduction of ten
dimensional SYM theory to the dual torus.

We would like to emphasize that the local dynamics of the Super Yang
Mills (SYM) theory on the dual torus, which encodes the dynamics of the
compactified matrix model, is not of direct physical relevance in the
matrix model.  Indeed, translations of the dual torus coordinate are
matrix model gauge transformations by the unitary ``matrix'' $e^{i{\bf
\alpha P}}$ (${\bf P}$ is the torus translation generator).  We will see
that on the subset of matrix model degrees of freedom which represent
strings, invariance under this gauge transformation becomes the Virasoro
condition $L_0 = \overline{L_0}$ of light cone gauge string theory.
States which do not satisfy this condition, {\it i.e.}\ states which
carry momentum in the SYM theory, will be interpreted as strings
stretched along the longitudinal direction.  The true 
dynamics of M theory corresponds to scattering of SYM excitations in the
moduli space of the SYM theory.

Another unusual feature of the SYM theory which arises from the matrix
model is that its coupling constant scales as $g^2_{SYM} \sim R_9^{-1}$,
(more generally it scales like the volume of the dual torus).  This is
because the integral over the dual torus coordinates arises as the limit
of the trace in the matrix model.  The trace of the unit matrix is the
total longitudinal momentum of the system, and should be independent of
$R_9$.  This is achieved by rescaling the coupling.  The appropriate
dimensions of the coupling are made up by powers of the eleven
dimensional Planck length, which we set equal to one.

\nref\witteno{E. Witten, {\it Theta Vacua in Two-dimensional Quantum
Chromodynamics}, Nuovo Cim. {\bf 51A}(1979)325.}%
\nref\witten{E. Witten, {\it Bound States of Strings and P-branes},
Nucl. Phys. {\bf B460}(1996)335, hep-th/9510135.}%

The gauge potential $A$ can be gauged away in one spatial dimension,
apart from its Wilson line degree of freedom.  As $R_9 \rightarrow 0$,
the radius of the $\sigma$ circle goes to infinity.  In this limit, the
quantum dynamics of the Wilson lines is frozen.  The field strength
variable conjugate to the Wilson line becomes a classical variable
${\cal E}$, which behaves like a two dimensional $\theta$ angle
\refs{\witteno, \witten}.
Its allowed values are discrete, corresponding to Casimir operators in
various representations of the gauge group (the lowest Casimir operator
in each conjugacy class).  For a given background electric field ${\cal
E}$, the energy scales like $R_9^{-2} {\cal E}^2$.  The first factor of
$R_9^{-1}$ comes from the volume of the dual circle and the second
factor of $R_9^{-1}$ from
the scaling of the coupling described in the previous paragraph. 

We will identify type IIA strings with the degrees of freedom in $U(1)$
subgroups of the $U(M)$ group\foot{It is important that we work in the
light cone gauge rather than in the static gauge.  In the light cone
gauge the two spinors on the world sheet of the IIA string have opposite
space time chirality.  This is exactly as we find in the $U(1)$ gauge
theory.  This is to be distinguished from the IIB theory, where they have
the same chirality.  In the standard study of D strings in IIB theory, 
the static gauge is used. There the space time chiralities are opposite.}.
If we rescale $\sigma$ to go from $0$ to $2\pi$ and $X$ and $t$ so that
the quadratic terms in the Hamiltonian are independent of $R_9$, then
the commutator terms scale like $R_9^{-3}$ (bosonic) and $R_9^{-3/2}$
(fermionic).  Thus, in the $R_9 \rightarrow 0$ limit we should restrict
attention to commuting matrices.
As shown in \bfss, the
matrix model Lagrangian reduces to the multiple copies of the Type IIA
Green-Schwarz lagrangian on this subset of matrix configurations.  Our
general comment about gauging of translations in the field theories
which represent compactifications of the matrix model, shows that the
correct Virasoro constraints of the light cone Green-Schwarz superstring
follow from the gauge symmetries of the matrix model\foot{This result
was shown independently in \Motl.}.

States of the field theory which do not satisfy the level matching
constraint can be viewed as strings wrapped around the longitudinal
direction.  This follows from the usual Virasoro equation of light cone
gauge field theory $\partial_{\sigma} X^- = p(\sigma )$, where $p$ is
the world sheet momentum density.   A similar phenomenon will arise in
Section 3, when we study the emergence of Type IIB strings from $\sym
2$.  There the values of world volume momenta 
 are the charges under the NS-NS and the
R-R two form gauge potentials for longitudinally wound elementary and
Dirichlet strings.  This  may be seen by examining the
supersymmetry algebra of the $2+1$ dimensional theory
\eqn\twoplussusy{\{ Q^A_\alpha, Q^B_\beta\} = 2 \delta_{\alpha \beta}
\gamma_i^{AB} p^i}
where $i=0,1,2$ and $A,B=1,2$ are indices of the $2+1$ Lorentz symmetry
(vector and spinor respectively) and $\alpha,\beta=1,...,8$ are
$spin(8)$ indices.  From the space time point of view \twoplussusy\ is
interpreted as part of the IIB supersymmetry algebra in the light cone
frame.  Here $\alpha,\beta$ are spinors of the transverse Lorentz
$spin(8)$ symmetry and $A,B$ label the two supercharges of the type IIB
theory.  The Hamiltonian $p^0$ in \twoplussusy\ is the minus component
of the space time momentum $P^-$.  $p^{1,2}$ appear precisely as the two
central charges for strings stretched along the longitudinal direction.
More explicitly, as in
\ref\bss{T. Banks, N. Seiberg and S. Shenker, {\it Branes from
Matrices}, hep-th/9612157}
we can identify them with
\eqn\centralchar{\eqalign{&p^1= \dot X^a [X^8, X^a] + ... = \dot 
X^a D_1 X^a + ... \cr
&p^2= \dot X^a [X^9, X^a] + ... = \dot X^a D_2 X^a + ... \cr}} 
where $a$ runs over the transverse directions and we used the fact that
$X^{8,9}$ become the two covariant derivatives in the spatial
directions. 

We wish to make one further comment on the construction presented in
\bfss .  It produces a chiral two dimensional field
theory as a limit of finite matrix constructions.  This is not terribly
surprising.  From the matrix model point of view, the derivative
operator arises as the limit of the matrix $p$, which is taken to be a
matrix with eigenvalues equal to the $1\over i$ times the logarithm of
the $N$'th roots of unity, and to commute with the matrix $V$ of \bfss.
Thus, our construction resembles the SLAC derivative of lattice gauge
theories 
\ref\slac{S.D. Drell, M. Weinstein, S. Yankielowicz, {\it Strong
Coupling Field Theories: 2. Fermions and Gauge Fields on a Lattice},
Phys. Rev. {\bf D14}(1976)1627.}.
In the matrix model, the lack of periodicity in the spectrum of $p$ is
required to describe wrapping configurations of membranes.

The strings which were exhibited in \bfss\ all have {\it the same}
longitudinal momentum.  Motl \Motl\ has described how strings with
larger values of $p_L$ emerge from the matrix model\foot{Motl's
construction was prefigured in work on black hole dynamics in string
theory 
\ref\malda{J. Maldacena, L.Susskind, {\it D-branes and Fat Black Holes},
Nucl. Phys. {\bf B475}(1996)679, hep-th/9604042; S. Das, S. Mathur,
{\it Excitations of D strings, Entropy and Duality}, Phys. Lett. {\bf
B375}(1996)103, hep-th/9601152.}.}.
We have argued above that the stable semiclassical configurations are
(in a particular gauge) diagonal $X^i (\sigma )$ matrices.  However,
there is no need for these matrices to be periodic in $\sigma$.  Rather,
they are periodic up to a gauge transformation, which preserves the
diagonal gauge, {\it i.e.}\ a permutation of eigenvalues.  $P\times P$
diagonal configurations, which have been ``screwed together'' in this
fashion by a permutation of rank $P$, are in one to one correspondence
with strings of $p_L = \epsilon P$, where $\epsilon$ is the unit of
longitudinal momentum carried by a $1 \times 1$ matrix.  
Furthermore, the prescription that the
light cone string coordinate measures the string's longitudinal momentum
follows immediately from this ansatz.  To get arbitrary ratios of $p_L$
we have to study the large $M$ limit.

It remains to show that the correct string interactions emerge from this
picture in the limit $R_9 \rightarrow 0$.  Motl has argued that the
correct scaling of the string coupling indeed emerges, but much work
remains to be done along these lines.  

Motl's construction has a beautiful interpretation in terms of the
moduli space of SYM theory.  We will show that this can be generalized
to describe arbitrary toroidally compactified IIA and IIB strings.  
We will therefore turn in the next section to a general description of
these SYM moduli spaces.  In Section 3 we show how the Coulomb branch of the
moduli space of toroidally compactified $\sym d$ maps, in the limit that
one radius of the SYM torus is much larger than others, into the Fock
space of light cone gauge IIA string field theory compactified on the
torus dual to the small SYM directions.  This embedding in SYM theory
provides a natural nonperturbative prescription for string interactions.
We then show how a similar picture for Type IIB strings
emerges in another limiting regime of the torus geometry, as first
proposed by Aspinwall and Schwarz 
\ref\aspsch{P. Aspinwall, {\it Some Relationships Between Dualities in
String Theory}, Talk Presented at ICTP Trieste Conf. on
Physical and Mathematical Implications of Mirror Symmetry in String
Theory, June 1995, hep-th/9508154; J. Schwarz, {\it The Power of M
Theory}, Phys. Lett.{\bf B367}(1996)97, hep-th/9510086.}.  

\newsec{\bf Some Properties of SYM Theories with Sixteen Supercharges}

The properties of SYM theories with sixteen hermitian supercharges have
recently been investigated in 
\ref\nati{N.Seiberg, {\it Notes on Theories with 16 Supercharges},
RU-97-7, to appear.}. 
Here we will present some relevant results of the analysis of these
theories, and refer the reader to \nati\ for more details.

\subsec{The $d=3$ Theory}

We start with the study 
of field theories with $N=8$ supersymmetry in $d=3$.  The super
generators are in the real two dimensional representation of the Lorentz
group.  The automorphism of the algebra (R symmetry) is $spin(8)_R$ and
the supergenerators transform as an eight dimensional representation,
which we take to be the spinor $\bf 8_s$.

Since for massless particles the little group is trivial, there is only
one massless representation of the superalgebra.  It consists of 8
bosons in the $\bf 8_v$ of the R symmetry and 8 fermions in the $\bf
8_c$.  Starting in a higher dimensional field theory with the same
number of supersymmetries (e.g.\ $N=4$ in $d=4$) we find a vector field,
7 scalars and 8 fermions.  The R symmetry, which is manifest in this
description, is $spin(7) \subset spin(8)_R$.  The vector is a
singlet of $spin(7)$, the scalars are in $\bf 7$ and the fermions in
$\bf 8$.  After performing a duality transformation on the vector it
becomes a scalar.

Interacting Lagrangians with $N=8$ supersymmetry do not necessarily
exhibit the maximal possible R symmetry.  In particular, the Yang-Mills
Lagrangian is invariant only under the $spin(7)$ subgroup.  

The gauge coupling $g$ has dimension $\half$, and therefore the theory
is superrenormalizable.  To analyze its long distance behavior we start
by considering the moduli space of vacua.  Along the flat directions the
$U(M)$ gauge symmetry is broken to $U(1)^M$.  The low energy degrees of
freedom are in $M$ identical free $N=8$ multiplets, each of which
includes seven scalars $X^{Ai}$ ($i=1,...,7$) and a photon.  
The dual of the photons are compact scalars $\phi^A$ which live on the
Cartan torus of $U(M)$ (in general, they live on the Cartan torus of the
dual gauge group \nati).  The Lagrangian is:
\eqn\duallag{{ 1\over g^2} (\partial X^{Ai})^2 + g^2 (\partial
\phi^A)^2.} 
Because of $N=8$ supersymmetry the Lagrangian \duallag\ is not corrected
in the quantum theory.  Therefore, the $8M$ real dimensional moduli space
of vacua $\CM$ is flat.  The $X^{Ai}$ label $\BR^{7M}$ and $\phi^A$
labels $\bT^M$.  The Weyl group of $U(M)$ is $\bS_M$.  It permutes the
$A$ indices and so  
\eqn\modsp{\CM={\BR^{7M}\times \bT^M \over \bS_M} .}
It has singularities whenever $X^{Ai} = X^{Bi}$ and $\phi^A = \phi^B$
for some $A$ and $B$ and all $i=1,...,7$.  The metric
around these singular points is an orbifold metric.

Since the theory is superrenormalizable, the only dynamics which
survives at energies smaller than $g^2$ is the infrared dynamics of
massless modes.  Note that if we define $g^2 \phi^A = X^{A8}$, to
emphasize the $spin(8)$ symmetry of the lagrangian, then the radius of
the $X^{A8}$ torus goes to infinity and we can focus on a neighborhood
in the moduli space.  At the generic point we find a free field theory.
The theory at the orbifold singularities is more interesting.  The
moduli space around each of them looks like $\BR^{8M}/ \bS_M$.  We
believe that the theory describing these points is an interacting
superconformal fixed point.  A more extensive discussion is given in
\nati.

At long distance, the theory must flow to a scale invariant theory.  It
is expected that if the theory is interacting, it is also superconformal
invariant.  The conformal algebra in 3 dimensions is $spin(3,2)$.  The
eight supersymmetry generators combine with eight superconformal
generators to eight spinors of $spin(3,2)$.  For the closure of the
algebra we must include the the $spin(8)_R$ symmetry
\ref\nahm{W.Nahm, {\it Supersymmetries and Their Representations}, Nucl.
Phys. {\bf B135}(1978)149.}.

The long distance theory is scale and superconformal invariant.  As such
it has a global $spin(8)$ symmetry which acts as an automorphism of the
supersymmetry algebra.  Along the flat directions the long distance
theory is free and then the $spin(8)$ symmetry is manifest.

Below we will interpret these results in terms of the derivation of Type
IIB string theory from the matrix model.  The emergence of the $spin(8)$
symmetry in the infrared dynamics of the $2+1$ dimensional theory will
imply that the string theory has an eight dimensional rotational
invariance relating a dimension which arises from membrane winding to
the manifest noncompact dimensions of the matrix model.  The fact that
the two $(2+1)$ Lorentz components of the eight SUSY generators transform
in the same spinor representation of $spin(8)_R$ will there imply that
the spacetime SUSY of the string theory is the chiral IIB algebra.
 
\subsec{Compactification from $d=4$}

Consider now starting in a higher dimensional theory with 16
supercharges and compactifying on a torus to three dimensions.  Some of
the scalars in the three dimensional Lagrangian originate from
components of gauge fields in the higher dimensional theory.  Therefore,
the corresponding directions in the moduli space of the three
dimensional theory must be compact.  Let us start by considering the
free $U(1)$ $N=4$ theory in $d=4$ with gauge coupling $g_4$ and
compactify it on a circle of radius $R$ to three dimensions.  The three
dimensional gauge coupling $g_3$ satisfies
\eqn\threedgac{ {1 \over g_3^2} = {R \over g_4^2}.}
The six scalars in the vector multiplet in four dimensions become
$\phi^i$ with $i=1,...,6$.  $\phi^7 $ arises from a component of the
four dimensional gauge field $\phi^7=A_4$.  It corresponds to a $U(1)$
Wilson line around the circle.  A gauge transformation, which winds
around this circle, identifies $\phi^7 $ with $\phi^7 + {1 \over R}$.
Therefore, we define the dimensionless field $\phi_e= RA_4$, whose
circumference is one.  When we dualize the three dimensional photon to a
scalar $\phi_m$, we find the Lagrangian 
\ref\swcompactification{N. Seiberg and E. Witten, {\it Gauge Dynamics
and Compactification to Three Dimensions}, hep-th/9607163.}
\eqn\fourthreelag{{R \over g_4^2}(\partial \phi^i)^2 +{1 \over
Rg_4^2}(\partial \phi_e)^2 + {g_4^2 \over R} (\partial \phi_m)^2.}
The moduli space of vacua is 
\eqn\fourtothree{\BR^6 \times \bT^2 }
where the two circles in $\bT^2$ correspond to the two compact bosons
$\phi_e$ and $\phi_m$.  They represent a $U(1)$ Wilson line and a $U(1)$
'tHooft line around the circle we compactified on. In other words, these
two scalars are the fourth component of the $d=4$ photon $A_4$ and the
fourth component of the magnetic photon $\tilde A_4$.  The non-trivial
duality transformation in $d=4$ is translated to
\eqn\fourddul{\eqalign{
\phi_e \rightarrow \phi_m \cr
\phi_m \rightarrow - \phi_e \cr
g_4 \rightarrow {1 \over g_4} .\cr}}
It is easy to add the $\theta$ angle in four dimensions and recover the
$SL(2,\BZ)$ action in four dimension as an action on the $\bT^2$ in the
moduli space \fourtothree\ \swcompactification. 

As we said above, at long distance in the three dimensional theory only
the local structure of the moduli space \fourtothree\ matters.  It is
$\BR^8$.  The eight scalars transform as a vector under the enhanced
$spin(8)_R$ symmetry.  The duality transformation \fourddul\ becomes
part of the $spin(8)_R$ symmetry.

We can easily extend this discussion to compactified interacting
theories.  For example, consider the $U(M)$ $N=4$ theory in $d=4$.
Repeating the analysis of the $U(1)$ theory and modding out by the Weyl
group, we find the moduli space of vacua
\eqn\fourtothreet{{\BR^{6M} \times \bT^{2M} \over {\bf S}_M.}}

The full theory is invariant only under the $spin(6)$ symmetry of the
four dimensional theory.  The $SL(2,\BZ)$ duality is {\it not} a
symmetry of the theory.  It relates theories with different values of
the coupling constant.  After the compactification this $SL(2,\BZ)$ acts
on the $\bT^{2M}$ factor (it acts as the usual discrete diffeomorphism
symmetry on each of the $M$ $\bT^2$ factors).  Again, it is not a
symmetry.  However, at long distance its $\BZ_2$ subgroup \fourddul\
becomes a symmetry.  Therefore, the symmetry at long distance includes
$spin(6)\times \BZ_2$.  The three dimensional Lagrangian is obtained by
shrinking the compactification radius $R$ with $g_3$ fixed.  Then, the
$spin(6)$ R symmetry of the four dimensional theory is enhanced to
$spin(7)$, which is manifest in the three dimensional Lagrangian.  Since
in this limit $g_4 \rightarrow 0$, the $\BZ_2$ subgroup of $SL(2,\BZ)$
is not visible. In the long distance limit we should find a symmetry,
which includes both this $spin(7)$ R symmetry and $spin(6)\times \BZ_2$.
This must be $spin(8)$.  This leads to an independent derivation of the
$spin(8)_R$ symmetry of the long distance theory (the other derivation
was based on its superconformal invariance).  This argument is similar
to that of \ss.

We conclude that the electric-magnetic duality of the four dimensional
theory becomes a symmetry of the three dimensional theory at long
distance.  It is included in its $spin(8)_R$ R symmetry.

\subsec{Generic Toroidal Compactifications of $\sym d$}

The final result which we will need in our discussion of the matrix
model is that for the moduli space of $\sym d$ compactified on a torus
of generic dimension.  The term generic means that we will omit
discussion of the special consequences of duality in low dimensions.
We will also restrict attention to the gauge group $U(M)$.

If we compactify $\sym d$ to $k$ noncompact spatial dimensions, we
obtain $\sym k$, which contains $9-k$ scalar fields in the adjoint
representation of $U(M)$.  $9-d$ of these fields, $X^i$, are noncompact
variables.  The other $d-k$ arise from integrating the $d$ dimensional
gauge potentials over one cycles of the $d - k$ torus.  We call these
variables $\Phi^a$.  Along the generic flat direction, the gauge group
is broken to $U(1)^M$.  The $M(d - k)$ variables $\Phi^a$ are now angle
variables which live in $(d - k)$ copies of the Cartan torus of $U(M)$.
The moduli space is thus
\eqn\modcompg{\BR^{M(9-d)} \times \bT^{M(d - k)} \over \bS_M}
The kinetic term for the compact fields takes the form
\eqn\anglag{ {1\over g_k^2} G_{ab} \partial\Phi^a \partial\Phi^b}
where $g_k^2$ is the effective $\sym k$ coupling, including a factor of
the inverse volume of the $d - k$ torus.  $G_{ab}$ is the metric of
$\bT^{M(d - k)}$.  Since the $\Phi^a$ are Wilson loops, integrals of
gauge field components along cycles of the original torus, the scale of
this torus is the inverse of the compactification size.  For example,
for $M=1$, $\bT^{d-k}$ is the dual of the compactification torus, while
for general $M$ it is the product of $M$ copies of this dual torus.  In
the matrix model application below, it is this dual torus which plays
the role of the spacetime on which strings propagate.

\newsec{M Theory on Tori}

\subsec{Generalities}

The compactified matrix model is $\sym d$.
Compactified IIA strings should be thought of as M theory $2$-branes
wrapped around a one dimensional cycle of $\bT^d$.  In the weakly
coupled type IIA theory from which the matrix model was extracted in
\bfss, the membrane is described as a Dirichlet brane\foot{The reader
should carefully distinguish the Type IIA string theory in the present
paragraph from that discussed in the rest of the paper.  Here,
the longitudinal direction is thought of as
small, while $\bT^d$ is of string scale.  IIA strings are membranes
wrapped around the longitudinal direction.  We will quickly return to a
situation in which the longitudinal direction is large, where we derive
another copy of perturbative IIA strings by taking a transverse
dimension small.}.  In the T dual
prescription which leads to $d + 1$ SYM theory, membranes wrapped around
a single cycle of the torus are, from the SYM point of view, $d-1$
dimensional domain walls wrapped around the dual cycle of the dual
torus.  So, a membrane wound around the ninth direction of a rectilinear
torus is, in SYM language, a domain wall wrapped around the first $d -
1$ directions (we label the dual torus directions by $\sigma^{a - 9 +
d}$ for the direction dual to the $a$th direction in target space).
Having identified these configurations in the weakly coupled Type IIA
limit we now go to strong coupling via the conjectures of \bfss.
Namely, the theory which describes the short distance interactions of
zero branes at weak coupling, is taken to be the entire theory at strong
coupling. What we have learned via this excursion is how to identify the
degrees of freedom which will represent Type IIA strings in another weak
coupling limit in which the longitudinal direction is large while one of
the transverse directions is shrunk to zero.  We will see that the
representation of IIA strings as $d - 1$ dimensional domain walls arises
naturally from $\sym d$ itself.

The limit of $\sym d$ which is supposed to describe IIA string theory
compactified on a torus is one in which the radius $R_9$ is taken very
much smaller than the eleven dimensional Planck scale, while the other
dimensions are taken large.  Indeed, the typical size of these other
directions are of order the scale set by the weakly coupled Type IIA
string tension.  From the SYM point of view this means that we have one
large and $d-1$ small dimensions, and it is clear that, to first
approximation, we should ignore modes which carry momentum in the small
directions.
Thus, directly in the SYM theory, we can understand that the degrees of
freedom which dominate the IIA limit are $1+1$ dimensional fields,
corresponding to integrals of the underlying degrees of freedom over $d
- 1$ dimensional domain walls.

Of course, what we have done here is to dimensionally reduce $U(M)$
$\sym d$ to $\sym 1$.  As we discussed in the previous section
the moduli space of the dimensionally reduced theory is $\BR^{M(9-d)}
\times \bT^{M(d - 1)}/ \bS_M $.

Dynamics along the moduli space is thus described by eight free
$1+1$ dimensional scalar fields and their superpartners, modded out by
a discrete gauge symmetry.  The boundary conditions obeyed by these
scalar fields may be twisted by any element of the discrete group, which
is the semidirect product of the weight lattice of $U(M)$ and its Weyl
group
\eqn\discretegroup{\IZ^M \sdtimes \bS_M.}

The conjugacy classes of this group are easily worked out.  Each group
element is the product of a permutation and a shift. Write the
permutation as a product of commuting cycles.  It is obvious that we can
work within the subspace corresponding to a given cycle.  The
permutation is then the cycle $S$ which takes $x_k \rightarrow x_{k+1}$.
Conjugating this by a shift $x_k \rightarrow x_k + s_k$ we get the
product of $S$ and a shift by $s_k - s_{k+1}$.  This fails to be a
general shift because it is traceless.  Thus the most general element is
conjugate to a permutation times a shift which shifts the whole subspace
acted on by each cycle of the permutation by the same lattice vector.

This is precisely what we need for the interpretation of the moduli
space as compactified Type IIA string theory. 
The sector corresponding to a given permutation $S$ is interpreted in
string theory language as follows: 
Writing $S$ as a product of commuting cycles, and arranging the matrix
elements of the $X^i$ in cyclic order, we obtain Motl's picture of long
strings.  The sector of the gauge theory with $S$ a product of $k$
cycles of lengths $p_k$ is in one to one correspondence with multistring
states of $p_k$ units of longitudinal momentum.  Note that the usual
light cone correspondence between string length and longitudinal
momentum follows directly from the matrix model identification of the
longitudinal momentum with the rank of matrices.  The shifts provide us
with the winding numbers of these strings around cycles of the torus.  
The nontrivial conjugacy classes correspond to assigning a winding
number around each cycle to each string of each value of the
longitudinal momentum.  In particular, we do not have separate winding
numbers for the different diagonal elements of the matrix which makes up
a long string.

We note that in the full $U(N)$ theory, 
the permutation sectors are not really topological, since 
permutations can be continuously deformed to the identity in $U(N)$.
However, it is easy to see that as $R_9 \rightarrow 0$, the masses of
the fields transforming as roots of the Lie algebra go to infinity.   
Combining this with SUSY nonrenormalization theorems we see that in this
limit the free string picture becomes exact.  The different sectors,
representing strings with different values of longitudinal momenta, do
not transform into each other.
The challenge of deriving
string interactions as corrections to this limit will be taken up elsewhere.  

We have thus shown that the large $M$ $\sym d$ theory, reproduces, in
the appropriate limits both compactified and uncompactified Type IIA
string theory.  We have worked in the limit $R_9$ goes to zero with
other radii fixed.  In this limit the strings are free.  We can thus do
the usual T duality transformations on them.  What is not clear at this
juncture is how the string interactions defined by $\sym d$ transform
under those transformations.  Since the string variables are only a
small subset of the $\sym d$ degrees of freedom, this is far from
obvious, particularly for those values of $d$ in which $\sym d$ is
nonrenormalizable.  Moreover, we do not expect the transformation rules
to be simple, since T duality in a single circle is supposed to
reproduce Type IIB string theory.

The nonperturbative formulation of Type IIB string theory in ten
dimensions is instead supposed to derive from the theory at hand by
taking a different limit of the radii.  We turn to this problem in the
next subsection.

\subsec{The Matrix Model on $\bT^2$ and Type IIB Strings}

As shown by Aspinwall and Schwarz \aspsch,
ten dimensional Type IIB string theory is supposed to emerge when M
theory is compactified on $\bT^2$ whose area goes to zero at fixed
complex structure.  The eleven dimensional Planck scale is rescaled so
that the $(p,q)$ string tensions $\sqrt{p^2 R_9^2 + q^2
R_8^2}l_{11}^{-3}$ are kept fixed.  Here we see this in the framework of
the matrix model.

The corresponding $\sym 2$ theory lives on a torus with area going to
infinity.  As we noted above, the SYM coupling scales as $g^2 \sim {(R_8
R_9)}^{- 1} l_{11}$ which means that it scales to infinity like $R^{-
5/3}$ (in string tension units) as we go to the Type IIB limit.  We
recall that $g^2$ has dimensions of mass in three dimensions.  Thus, in
three space time dimensions, the Yang Mills coupling is relevant.  SYM
theory with sixteen SUSY generators has a classical moduli space which
(apart from singular points) consists of abelian field configurations.
Near the singular points of the moduli space the theory is likely to be
described by a nontrivial infrared fixed point.  The scalings noted
above suggest that apart from this extreme infrared dynamics on the
moduli space, all other features of $\sym 2$ will decouple from the
dynamics in the IIB limit.  The limiting theory will be described by
infrared fixed points, trivial along the flat directions in the moduli
space and perhaps nontrivial near the singularities.

As we discussed above, the strong coupling limit of $\sym 2$ has a
$spin(8)$ global symmetry.  In terms of the membranes it rotates the
space time momentum component which arises as membrane winding number
into the ordinary transverse space time momenta.  It is interesting how
the required enhanced Lorentz symmetry, which should arise in the area
going to zero limit, appears in the strong coupling limit of the $\sym
2$ theory.

To actually see the IIB strings in $\sym 2$, we must integrate the
moduli fields over a one cycle of the dual torus.  This follows the
general prescription we have described above for finding strings in
$\sym d$.  The new wrinkle here is that the appropriate fields to
integrate include the {\it dual} to the photon field, rather than the
gauge field itself.  This is because we have a strongly coupled gauge
theory in the infinite coupling limit.  Then the dynamics is fully
described by the moduli.  In order to see the proper scalings of the
Lagrangian, we will redo the duality transformation.

In terms of a gauge coupling $g^2$
and a flat metric $G_{\mu\nu}$, the duality transformation of a three
dimensional gauge theory is performed by doing the functional integral
over $F_{\mu\nu}$ of the action,
\eqn\threeddual{S = \int ({\sqrt{G}\over g^2}
[G^{\mu\nu}G^{\lambda\kappa}F_{\mu\lambda}F_{\nu\kappa} +
G^{\mu\nu}\partial_{\mu} \phi^i \partial_{\nu} \phi^i ] 
+ \epsilon^{\mu\nu\lambda}F_{\mu\nu}\partial_{\lambda}\phi^8 ),}
where $i$ runs from one to seven.  The integral leads to 
\eqn\dualact{S = \int [{\sqrt{G}\over g^2}^{-1} G^{\mu\nu}\partial_{\mu}
\phi^8 \partial_{\nu} \phi^8
 +{\sqrt{G}\over g^2}G^{\mu\nu}\partial_{\mu} \phi^i \partial_{\nu}
 \phi^i]}
In the present context, ${\sqrt{G}\over g^2}$ is independent of the
radii of the torus, and the $spin(8)$ invariance is manifest.  The
nonvanishing metric components are $G_{00} = 1$,$G_{11} = 1/R_8^2$ and
$G_{22} = 1/R_9^2$, all in eleven dimensional Planck units.

Elementary IIB strings are found by taking $R_9 \ll R_8$.  In this
limit, fields which vary in the $\sigma^2$ direction on the dual torus
are energetically costly.  The low energy excitations are functions only
of $\sigma^1$.  Thus, as in the case of compactified IIA strings, the $1
+ 1$ dimensional nature of the low energy excitations is a consequence
of taking one cycle of a (spacetime) torus much smaller than the others.
It is easy to verify that on this set of configurations, the $\sym 2$
moduli space Lagrangian reduces to multiple copies of the Green Schwarz
IIB Lagrangian:
\eqn\twobact{\int dtd\sigma^1 [(\partial_0 X^A )^2 + R_9^2
(\partial_{\sigma^1} X^A )^2] + \quad {\rm fermions}}
Note that unlike the construction of the IIA theory, here the two
spinors on the world sheet have the same space time chirality.  This is
exactly as it should be on the world sheet of the IIB string in the
light cone gauge.

Our analysis of the correspondence between sectors in the moduli space
Lagrangian and the Fock space of strings with arbitrary longitudinal
momentum goes through as well.  The only subtle point is that $X^8$ is a
periodic variable, but its period goes to infinity in the zero area
limit.

We can also describe Dirichlet strings in this formalism.
 We simply perform an
$SL(2,\IZ)$ transformation on the elementary string.  The Lagrangian is
not invariant under this.  The metric $G_{ij}$ transforms as $G
\rightarrow M^T G M$ where $M$ is the $SL(2,\IZ)$ matrix
$\pmatrix{m&n\cr p&q}$.  This reproduces the correct formula for the
Dirichlet string tensions \aspsch.  Of course, closed D-strings in ten
dimensions are not stable excitations.  They interact strongly and will
decay rapidly.  We do not yet know how to derive these interactions from
the matrix model.  Despite these {\it caveats} our derivation of the
Dirichlet string tensions is a correct one because we can apply it to
large smooth string configurations which approach the infinite straight
BPS strings.

We have given only a brief description of IIB strings here, since
everything follows precisely the pattern outlined
 by Aspinwall and
Schwarz \aspsch.  Nonetheless it is rewarding to see it emerge so nicely
{}from the matrix model formalism.

The domain wall character of the IIB string excitations of $\sym 2$
removes what might have been a paradox in the emergence of IIB strings
as a zero area limit.  Formally, the theory on the dual torus goes to
infinite volume in this limit and we might have imagined that the
rotation group in the toroidal volume is restored (and perhaps even
elevated to the $spin(2,1)$ Lorentz group).  However, there is no such
restoration of symmetry for the wrapped excitations which we are
studying. These always feel the toroidal boundary conditions.  The
question of the significance of local excitations of the SYM theory (for
which the restored symmetry might have some significance) deserves
further study.  We note in addition that the discrete subgroup of
rotations which preserves the toroidal boundary conditions when $R_8 =
R_9$ is clearly a gauge transformation of the matrix model.  It is
induced by a unitary transformation of the fundamental matrices which
preserves the trace in the compactified theory.  In the present case it
is simply the $\tau \rightarrow - {1\over \tau}$ subgroup of the
$SL(2,\IZ)$ gauge symmetry of toroidally compactified M theory. 

It is of some interest to understand more completely the role of the
$spin(2,1)$ Lorentz group and its extension to the conformal group at
the nontrivial fixed point.  It is clear that there can be no physical
symmetry between the time of the $2 + 1$ field theory, which is the same
as light cone time in the ambient spacetime, and its spatial dimensions,
whose corresponding translation generators are set equal to zero on
physical states.  Nonetheless, recalling the role of the light cone
Virasoro algebra in string theory, we may anticipate that these world
volume generators are crucial to the proof of ten dimensional Lorentz
invariance in the nonperturbative formulation of IIB string theory.  A
similar conclusion is also suggested by the connection which we pointed
out above, between states of nonzero two momentum and longitudinally
wrapped strings.

\newsec{\bf Conclusions}

We have shown that the prescription of compactifying the matrix model of
M theory as $\sym d$ on a dual torus, correctly reproduces the expected
string theories in various limits.  Our most complete results were for
ten dimensional Type IIA string theory.  The moduli space of large $M$
$\sym 1$ precisely reproduces the Fock space of light cone Type IIA
string field theory, and the SYM theory gives a nonperturbative
prescription for string interactions.  We have not yet shown that the
interactions it prescribes reduce to conventional perturbative string
theory in the zero radius limit.  The light cone level matching
condition follows from gauge symmetries of the matrix model which go
beyond those of SYM.  We also showed that toroidally compactified IIA
strings arise in the requisite manner from $\sym d$ .  Here our analysis
must be deemed less complete, if only because it does not really
distinguish those cases where $\sym d$ is a sensible continuum field
theory from those where it isn't.

Next we showed that the zero area limit of the compactification of the
matrix model on a two torus contained excitations which propagate like
free ten dimensional IIB strings with arbitrary $(p,q)$ charge (more
precisely, infinitely long strings carry charge, while the finite
excitations we have constructed do not).  In the limit of large complex
structure of the small torus, the $(0,1)$ string is weakly coupled and
even closed strings are almost stable.  The freely propagating strings
have a $spin(8)$ symmetry rotating the membrane winding number direction
into the ordinary dimensions of space. We gave an argument based on
$2+1$ dimensional field theory that this is an exact symmetry of the
model in the zero area limit.  We may anticipate that the discussion of
interactions will be more complicated in the IIB case, since it seems to
involve the construction of a nontrivial fixed point theory at the
origin of moduli space.

One of the most intriguing aspects of our study is the way in which the
$1 + 1$ dimensional character of string theory arises.  Weakly coupled
limits of toroidally compactified M theory arise when a single cycle of
the torus is much smaller than all the others.  In the $\sym d$
description this corresponds to $1$ large and $d -1$ small cycles and
leads to a Kaluza Klein reduction of the degrees of freedom which
accounts for the stringiness of the dynamics.  We are led to the
conclusion that in general, at strong coupling M theory is not stringy.
Rather, the picture which appears to emerge is that the dynamics on the
moduli spaces of supersymmetric field theories of higher dimension (we
emphasize that it is the moduli spaces which are to be thought of as
space time), is generally involved.  We anticipate a particularly
important role for superconformal fixed point theories, such as that
which we conjecture to describe the nonperturbative interactions of IIB
strings.

We cannot refrain at this point from making some remarks about the fact
that for $d> 3$ compactification on a $d$ torus leads to
nonrenormalizable field theories.  First we emphasize that as far as
spacetime is concerned, this is an infrared problem.  This follows from
the dual relation between the world volume of $\sym d$ and the
spacetime torus.  Shenker
\ref\steve{S.Shenker, {\it Private Communication}.} 
has also emphasized the way in which the ultraviolet $\sym d$ behavior
mirrors the infrared properties of the transverse Coulomb potential.
Thus, resolving this problem may tell us interesting things about low
energy spacetime physics.

There are a variety of possible responses to the problem of
nonrenormalizability.  The first is to search for a continuum definition
of $\sym d$.  For example, strongly coupled $\sym 4$ may be viewed as a
limit of $\bS^1$ compactification of a nontrivial fixed point theory in
$5+1$ dimensions.  Recent work of Rozali 
\ref\rozali{M.Rozali, {\it Matrix Theory and U Duality in Seven
Dimensions}, hep-th/9702136.} 
suggests that such a limit may indeed be relevant to the physics of the
matrix model compactified on a four torus.  We suggest that this may be
part of a larger story, and that once we understand compactifications
with fewer supersymmetries, a whole range of nontrivial fixed point
theories may prove to be relevant to the exploration of interesting
nonperturbative physics in M theory.  This would be analogous to the
role that $1 + 1$ dimensional conformal field theories play in
perturbative string theory.

With the full complement of SUSYs however, there does not seem to be a
possibility of nontrivial fixed points above $d = 4$ (note that even for
$d=4$ we have to appeal to toroidal compactification of a theory with
chiral SUSY in a higher dimension).  Perhaps this is related to the fact
that M theory only has two branes and five branes, as a consequence of
which nothing interesting happens when cycles of higher dimension shrink
to zero volume.  To make this remark more transparent, imagine defining
toroidally compactified $\sym d$ as the limit of a cut off
theory\foot{The most natural cutoff is one in which the derivatives on
the world volume are written as the limit of large $N$ matrices,
following one of the derivations of the $\sym d$ prescription from the
matrix quantum mechanics of \bfss. This cutoff preserves SUSY and
gauge invariance, and may be applicable to numerical approximations of
chiral and SUSY gauge theories in a more general context.  This will be
discussed in a future paper by one of 
the authors 
\ref\tb{T. Banks, {\it Manuscript in preparation}.}}.
If there are no strong coupling fixed points, the bulk dynamics of $\sym
d$ approaches that of free field theory as the cutoff is taken to
infinity.  However, in the toroidally compactified theory there are zero
modes whose infrared dynamics exhibits the full complications of lower
dimensional Yang Mills theory.  

Let us now remember that the relevance of the bulk $\sym d$ dynamics to
the physics of M theory is only apparent when we take a limit in which
all of the radii of the spacetime torus are taken much smaller than the
Planck length.  In other limits of the space of compactifications, the
$\sym d$ torus has fewer large dimensions and only the lowest momentum
modes around the small dual tori are included in the low energy
dynamics.   Thus, the {\it triviality} of high dimensional $\sym d$ may
be simply telling us that there are no interesting limits of the 
space of compactifications of the matrix model with unbroken eleven
dimensional SUSY, in which cycles of dimension higher than four are
shrunk to zero.  

We would like to stress an assumption that we have made implicitly
throughout this paper.  When considering situations in which one radius
of a torus was much larger than others, we have made the assumption that
we could do the standard dimensional reduction of $\sym d$ (or in the
IIB limit, of its dual theory).  While this is obviously correct along
the flat part of the moduli space, it remains an assumption about the
dynamics of whatever definition of the nonrenormalizable $\sym d$ theory
we may supply for $d > 3$.  We believe that this property is necessary
to the consistency of the rules for compactification of the matrix model
to various dimensions.  Decompactification of a spacetime circle should
always lead back to the theory compactified on one fewer dimension.

Finally, we would like to stress one of the main conclusions of 
the present work. The matrix model system, made up only of zero branes
and minimally stretched strings, is capable of reproducing the full
spectrum of various string theories in appropriate limits.  And it does
so within the context of a nonperturbatively defined, unitary theory of
string interactions.  In the various weakly coupled limits it is clear
that there are no low energy excitations apart from the appropriate
strings.  Thus the matrix model is sure to lead to an effective theory
of perturbatively interacting strings.  It remains to be seen whether
the interactions it prescribes are those of conventional string theory,
which are the only string interactions compatible with ten dimensional Lorentz
invariance.  This question is under active study
\ref\bssea{T. Banks, S.Shenker, L.Susskind {\it et. al.} ,{\it Work in
Progress}.}  
and we hope to be able to answer it soon.

\centerline{\bf Acknowledgments}\nobreak
This work was supported in part by DOE grant DE-FG02-96ER40559.  We
thank S. Shenker and L. Susskind for discussions.

\listrefs

\end